\documentclass[runningheads]{llncs}
\usepackage[T1]{fontenc}
\usepackage{graphicx}
\usepackage{array} 
\usepackage{booktabs} 
\usepackage[font=small]{caption}
\usepackage{float}
\usepackage{appendix}

\begin{document}
%
\title{MoodLoopGP: Generating Emotion-Conditioned Loop Tablature Music with Multi-Granular Features}

%
\titlerunning{Generating Emotion-Conditioned Loop Music with Multi-Granular Features}
%


\author{Wenqian Cui\thanks{Corresponding author} \and
Pedro Sarmento \and
Mathieu Barthet}

\authorrunning{Cui, Sarmento, Barthet}
%

\institute{Queen Mary University of London, School of Electronic Engineering and Computer Science, 327 Mile End Rd, Bethnal Green, London E1 4NS, United Kingdom \\
\email{cuiwenqian.app@gmail.com, p.p.sarmento@qmul.ac.uk, m.barthet@qmul.ac.uk}}

%
\maketitle              
%


\begin{abstract}
Loopable music generation systems enable diverse applications, but they often lack controllability and customization capabilities. We argue that enhancing controllability can enrich these models, with emotional expression being a crucial aspect for both creators and listeners. Hence, building upon LooperGP, a loopable tablature generation model, this paper explores endowing systems with control over conveyed emotions. To enable such conditional generation, we propose integrating musical knowledge by utilizing multi-granular semantic and musical features during model training and inference. Specifically, we incorporate song-level features (Emotion Labels, Tempo, and Mode) and bar-level features (Tonal Tension) together to guide emotional expression. Through algorithmic and human evaluations, we demonstrate the approach's effectiveness in producing music conveying two contrasting target emotions, happiness and sadness. An ablation study is also conducted to clarify the contributing factors behind our approach's results.

\end{abstract}
\section{Introduction}
\vspace{-2mm}

The significance of repetitive, loopable aspects in music structures is evident, especially in loop-centric genres like electronic dance music \cite{symbolicloop1}. Prior works have explored loop generation in both symbolic \cite{adkins2023loopergp,symbolicloop1,symbolicloop2} and audio domains \cite{audioloop1,audioloop2}, with some having specific focuses, such as drum instruments \cite{drumloop2,drumloop1}. However, increasing the degree of control in loop-based music generation systems is needed to address creative requirements, with agency over the emotions conveyed by the music standing out due to their direct inﬂuence on the listener’s experience and engagement. Emotion-controllable music offers potential applications in live performances, soundtracks, gaming \cite{affectivegaming2,affectivegaming1}, virtual/augmented reality (VR/AR), and even in personalized music generation and the data-driven musification in the context of smart cities \cite{musicalsmartcity}.

We utilize LooperGP \cite{adkins2023loopergp}, an advanced loopable symbolic music generation system that can effectively produce coherent and original loops with specified lengths, keys and time signatures, as our baseline. It extracts repeatable sections in music using a correlative matrix approach to derive the training data. This symbolic tablature generation system is trained on the DadaGP dataset \cite{sarmento2021dadagp}---a large-scale compilation of Guitar Pro format tablatures combining musical notes with playing techniques, dramatically elevating the expressiveness in the generated music. Such expressiveness can be harnessed for better emotional representation in music.

To guide our model in generating music conveying specific emotions, we add control tokens to the start of the symbolic token sequences, inspired by the GTR-CTRL model \cite{gtrctrl}. Our study mainly targets happiness and sadness, which are associated to two quadrants in the two-dimensional valence/arousal space based on Russell's model of affect \cite{russell1980circumplex}. Happiness and sadness are representative emotions from the high valence and high arousal quadrant (first quadrant) and the low valence and low arousal quadrant (third quadrant), respectively. Hence, our system operates under the assumption that music with high valence and high arousal expresses happiness, while music with low valence and low arousal expresses sadness. However, we acknowledge the bias of these assumptions and recognize that they may not hold in all contexts.

Even though earlier works have used valence and arousal scores as controls \cite{continuous,valencearousalMathieu}, we posit that it might not fully harness the model's potential for conditional generation. Motivated by the findings in psychological research \cite{tempomode,rhythmic_unit,tempomodetexture}, which explored the intrinsic musical features contributing to conveying distinct emotions, we integrate specific musical elements, notably tempo and mode, during both training and inference to enhance conditional generation capabilities. This is to investigate if the features highlighted by music psychology studies can also be advantageous to AI generative systems, and we note that the approach is not bounded by happy and sad emotions, for it can be extended to other emotions by leveraging correlated musical features.


While many features are significantly associated to music's emotional expression, they often remain static throughout a piece. Given music's dynamic nature, representing its essence with a single attribute is limiting. To address this, we introduce an approach integrating multi-granular features at both song and bar levels for emotion-conditioned generation. Specifically, we utilize tonal tension---metrics capturing tonal attributes---as bar-level features, based on their known correlation with musical emotions \cite{tonalemotion1,tonalemotion2}.

We trained our model on DadaGP \cite{sarmento2021dadagp}, a dataset specializing in Guitar Pro format guitar tablatures, with an encoder/decoder framework to convert symbolic tokens into Guitar Pro files. The Transformer-XL \cite{transformerxl} model is employed for sequence generation. Our results highlight the significance of both song and bar-level features in emotion-conditioned music, validated through algorithmic evaluations and a listening test. To summarize our contributions: 
1) We improved on LooperGP, a generation system that creates loopable music, by incorporating a control for emotion; 2) We incorporated features from music psychology research in the emotion control process alongside emotion labels; 3) We investigated enriching the emotional control process by integrating both song-level and bar-level features.


\vspace{-2mm}
\section{Related Work}
\vspace{-2mm}
\subsection{Emotion-Conditioned Symbolic Music Generation}
\vspace{-1mm}
To generate symbolic music with specific emotions, one common method is to insert emotion control tokens at the start of the sequence as conditions \cite{valencearousalMathieu}. This conditioning method is widely used in various tasks or domains. Sarmento et al. \cite{gtrctrl} use tokens to condition the instruments and genres of generated music, and Keskar et al. \cite{ctrltransformer} use control tokens to generate sentences with target attributes.

For other conditioning methods, Tan et al. \cite{fadernets} use low-level musical features to infer high-level features to perform music style transfer. Ferreira et al. \cite{genetic} uses genetic algorithms to condition mLSTM to generate video game soundtracks with certain emotions. Huang et al. \cite{tile} use the tile function to condition the CVAE-GAN architecture. Grekow et al. \cite{sampling} generate music with certain emotions by random sampling the 20-dimension latent space of CVAE. Instead of using discretized values, Sulun et al. \cite{continuous} use continuous-valued valence/arousal scores to condition a transformer to generate music, which is classified as the dimensional approach in \cite{emotionoverview}.

Emotions can also be inferred from other modalities. Tan et al. \cite{image1} uses image-music pairs with the same emotion to train and condition the music generation model, and Madhok et al. \cite{image2} uses the emotion vector classified using image to condition the music generation model.

\vspace{-2mm}
\subsection{Emotion-related Features in Music}
\vspace{-1mm}

The emotions perceived or felt upon listening to music have been extensively studied in literature, with researchers focusing on intrinsic features such as tempo and mode \cite{tempomode,tempomodetexture}. Dalla et al. \cite{tempomode} designed an experiment where the infants were asked to point to happy or sad faces after listening to music, and they found that fast tempo and major mode are related to happy music, while slow tempo and minor mode are related to sad music\footnote{Only major and minor modes were considered in this study.}. Webster et al. \cite{tempomodetexture} further investigated the combined effect of tempo, mode, and texture, showing that fast tempo, major mode, and simpler melodies result in happier music, while slow tempo, minor mode, and thicker texture result in sadder music.

Juslin et al. \cite{cue} examined how five acoustic cues regarding tempo, energy, and articulation, relate to the emotions of happiness, sadness, anger, and fear. Blood et al. \cite{PET} uses positron emission tomography (PET) to measure the relationship between musical emotions and the level of musical dissonance. Fernández-Soto et al. \cite{rhythmic_unit} investigates the tempo and rhythmic unit to four emotional semantic scales. Yang et al. (2023) \cite{multimodal_perception} highlighted that music emotion perception was a multimodal phenomenon that depended on less frequently studied features such as musical structure, performer expression, and stage setting, and was affected by individual factors such as musical expertise.

\vspace{-2mm}
\subsection{Tonal Tension}
\vspace{-1mm}

Tonal tension, as described in \cite{tonaltension}, quantifies the emotional and mental fluctuations induced by tonality in music. It is derived from the spiral array theory, representing pitch classes, chords, and keys in a helical three-dimensional space \cite{spiralarray}.
Tonal tension comprises three elements: cloud diameter, cloud momentum, and tensile strain \cite{tonaltension}. Cloud diameter gauges the maximal distance between any two notes within a cloud, while cloud momentum represents the distance between the centres of effect of two clouds of points, and tensile strain is the tonal distance between the centres of effect of a cloud of notes and the key. These metrics effectively quantify the tonality of a piece, and as such, are suggested as useful control tokens for emotion-conditioned music generation. By utilizing the varying values of tonal tension, more nuanced guidance is expected to be provided in the music generation process.

\vspace{-2mm}
\subsection{DadaGP and Guitar Tablature Generation}
\vspace{-1mm}
DadaGP \cite{sarmento2021dadagp} is a symbolic music generation dataset comprising 26181 guitar tablatures. It also contains an encoder/decoder to transform the guitar tablatures into symbolic tokens, which can be directly used to train sequence-to-sequence models. DadaGP covers 739 musical genres with a main focus on rock, metal, and their sub-genres.

DadaGP serves as a dataset for the generation of guitar and other instruments' parts in a tablature format. There are many works focus on guitar tablature generation, with most of them targeting a specific application. Sarmento et al. \cite{sarmento2021dadagp} trained a Transformer-XL model on the DadaGP dataset to generate guitar music in tablature. Sarmento et al. \cite{shredgp} focus on mimicking the style of four iconic guitarists by analyzing features from DadaGP. Loth et al. \cite{proggp} trained on a subset of DadaGP to generate progressive metal music. McVicar et al. \cite{guitarXML} focuses on generating guitar solo tablatures using MusicXML data.

\vspace{-2mm}
\section{Methodology}
\vspace{-2mm}
In this paper, we aim to enhance emotion-conditioned music generation by utilizing both song-level and bar-level features. We adopt Russell’s model of affect \cite{russell1980circumplex}, associating emotions to valence and arousal values in a two-dimensional space, to determine the level of happiness and sadness expressed by a piece. According to the model, high valence and arousal values correspond to happy emotion, while low valence and arousal values correspond to sad emotion. Therefore, we made an assumption that music with higher valence and arousal values is more likely to convey happy emotion, and vice versa.


To make the model generate music with target emotions, control tokens are added at the start of the token sequence. While only using valence and arousal is intuitive, we aim to explore the benefits of integrating other features. In this work, we classify the features into three categories: emotion labels and music psychology features as song-level features, and tonal tension as bar-level features.

In the following subsections, we will illustrate how we obtained the feature labels and incorporated the music loop information. Finally, we will summarize the pipeline of this work, covering data preparation, model training, and model inference.

\vspace{-2mm}
\subsection{Emotion Labels}
\vspace{-1mm}
To get the emotion labels for each song in the DadaGP dataset, we query the Spotify Web API using the artist name and the song title to retrieve the valence and energy values, where energy here serves as a surrogate to arousal, as in \cite{spotify}. The matching was carried out using the SpotiPy Python library. As a result of the matching process, a total of 16,173 songs were successfully annotated. 
The values obtained for valence and energy are continuous, but to use them as control tokens for the generative system, they must be discretized. This involves dividing them into two categories - high and low values. To determine the threshold for this division, the median valence and arousal values of all the pieces in the dataset are calculated and used. For instance, valence-high and valence-low are the tokens used for valence. Based on this categorization, music with high valence and high arousal is classified as happy music, while music with low valence and low arousal is considered sad music. In our case, considering ranges between 0 and 1, the thresholds for valence and arousal are 0.433 and 0.846, respectively.

\vspace{-2mm}
\subsection{Music Psychology Features}
\vspace{-1mm}
In this work, we focus on two features studied in the music psychology literature---tempo and mode. The mode of the music can also be found using the Spotify web API, and it is split into two classes: major mode and minor mode. Although tempo can also be found through Spotify web API, it is not extracted in this work because every song in DadaGP already has a token representing its tempo, and it is the necessary information for the decoder.


\vspace{-2mm}
\subsection{Bar-level Features}
\vspace{-1mm}

We utilize tonal tension (i.e., cloud diameter, cloud momentum, and tensile strain) as the bar-level features, employing the midi-miner package \cite{midiminor} for feature calculation from musical scores. Similarly, we discretized those values to use them as control tokens. We discretized bar-level features into four levels, using the first quartile, median, and third quartile of the data distribution as separating thresholds. The reason we use four levels instead of two levels to represent bar-level features is to support more combinations. Since the bar-level features are added for each bar of the music, they represent the ``state'' of that bar, and their purpose is to guide the music generation process. Hence, we would want the number of possible combinations of the features to be relatively large, so that they can represent more creative possibilities. In this study, we used three bar-level features. If these features had two levels, there would be a total of 8 possible combinations. However, if the features had four levels, the number of combinations increases to 64.

Since values can be derived for each bar of the music, we chose to append those values at the start of each bar of the piece, right after the \texttt{new\_measure} token) token. Therefore, the resulting sequence for every bar is: \texttt{new\_measure}, \texttt{cloud\_diameter}, \texttt{cloud\_momentum}, \texttt{tensile\_strain}, and then the rest of the tokens in this bar.

\vspace{-2mm}
\subsection{Keeping the Loop Information}
\vspace{-1mm}
The aim of this work is to generate emotion-conditioned and loopable music. Therefore, another part that should be integrated is to make the model generate coherent musical loops. Inspired by LooperGP \cite{adkins2023loopergp}, we use the same loop extraction method \cite{loopextraction} and the ``Barred Repeats'' method proposed in the paper, as it is shown the best result in the LooperGP paper.

\vspace{-2mm}
\subsection{The Overall Pipeline}
\vspace{-1mm}
In this section, we delineate our project pipeline, comprising data pre-processing, model training, and inference.

First, we query the Spotify web API to get the song-level features for every piece, including valence, arousal, and mode. We then perform the correlative matrix approach and the ``Barred Repeats'' method in LooperGP \cite{loopextraction} to get the loops used to train the model. The whole process of the loop extraction results in a further shrink of the dataset size to 13,466. Bar-level features, i.e., tonal tension, are then derived using the midi-miner package \cite{midiminor}.

After getting all the necessary features, the next step is to prepare the dataset for training. In the training process, every piece of music is represented by a token sequence. After discretizing all the features above, we add the control tokens to the corresponding positions within the sequence. We put the emotion labels and the music psychology features at the very beginning of the sequence, and put the tonal tension values right after every \texttt{new\_measure}\footnote{\texttt{new\_measure} is the token representing the start of a new bar.} token.

A Transformer-XL model \cite{transformerxl} is employed for the symbolic music generation task, predicting the next token in a sequence. Then, during inference, we use the control tokens to serve as a prompt to steer the model to generate music. Specifically, we want the model to be able to generate happy and sad music, so we use different prompts to make the model generate music with different emotions. We use the prompt sequence \texttt{[valence:high, arousal:high, mode:major, time\_signature:4]} to generate happy music, and use the prompt sequence \texttt{[valence:low, arousal:low, mode:minor, time\_signature:4]} to generate sad music. The thresholds for tempo are determined heuristically. We set an upper threshold of 150 BPM and a lower threshold of 100 BPM during inference, and sample the generated tempo to be higher or equal to 150 BPM for happy music and sample the tempo to be lower or equal to 100 BPM to generate sad music. Moreover, we allow the model to freely generate tonal tension without specifying values, assuming it learns to utilize them to guide the generation during inference.

Moreover, a time signature token is added during training. Although this was intended to ensure consistent metre, the model occasionally generates music with varying time signatures. Hence, post-processing steps are employed to regularize the output to 4/4 metres.

\vspace{-2mm}

\section{Experiments}

\vspace{-2mm}

As mentioned in the previous sections, we use both the song-level and bar-level features as control tokens to train the model and then use them as prompts in the inference process. We also conducted an ablation study to determine the contributing factors to the system.

The experiment settings include the following: the Transformer-XL model serves as the backbone of the symbolic music generation task, and we perform the next-token prediction task with cross-entropy loss. We trained each model for 100 epochs, with batch size being 8, learning rate being 0.0002, and AdamW as the optimizer.


\vspace{-2mm}

\section{Evaluations}

\vspace{-2mm}

In this section, we discuss the evaluation methods used in this work. This includes algorithmic approaches and a human-involved approach. The evaluation mainly focuses on the two essential aspects of this project, which are emotion and loop. Therefore, there are three main methods in our evaluation system, including training a neural network model to classify the emotions of the generations, a loop extraction algorithm to determine the number of loops in the generated music, 
and a subjective listening test to get human feedback on the generations in terms of loops and emotions.

All the evaluation processes are based on the model generations from the epoch 20 checkpoint. This is carefully chosen by ourselves to balance the music quality and the model's ability to generate emotion-specific music. It is mainly based on music quality and variability since the most important aspect of music generation is the music itself.
We found out that generations from earlier epochs would result in poor music quality, since the model has not learned the general music composition rules, and the generations from later epochs would result in serious overfitting of the training set since the variability of the model is poor and most of the generations are memorized from the training set. 
Based on the above criteria, we choose the final checkpoint from epoch 20. 

Different methods are evaluated for happy/sad music generated from the trained Transformer-XL model. During inference, the model generates 1000 pieces of happy music and 1000 pieces of sad music, and the 2000 pieces of music are evaluated using the algorithmic approaches. A Type I error $\alpha$ of 5\% is used in the statistical analyses.

\vspace{-2mm}

\subsection{Emotion Identification}

\vspace{-1mm}

We utilize the evaluation approach proposed in GTR-CTRL \cite{gtrctrl}, which is to train a neural network model for classifying the emotions expressed by the generated music. Based on GTR-CTRL, this BERT-style classifier effectively categorizes the attributes of the generated music from the symbolic tokens. We expand on the concept of using language classifiers for evaluation and extend it to include emotion classification.

We trained separate models for valence and arousal to measure the level of happiness and sadness in each piece, with both models sharing the exact same GPBERT architecture. We also use the median scores to discretize valence and arousal into binary classification labels. We trained the models on two parts of the data, including the original DadaGP data and the processed DadaGP data containing only the loops. We believe the loop subset of DadaGP might be a biased data source because it only contains parts of the music, whereas the valence and arousal scores by Spotify are derived from the entire piece of music, not just the loop part.

The training configurations are also the same as GTR-CTRL, which includes 768 tokens per song and the GPBERT layer, self-attention layer, feed-forward layer as the model architecture. We trained the models for 10 epochs and chose the best-epoch checkpoint for inference. The best result was achieved at epoch 6 for valence with a 70.89\% accuracy and epoch 2 for arousal with an 81.21\% accuracy. This result shows that the GPBERT model is slightly better at classifying arousal than valence.

We then use the trained models to classify the generated symbolic music. In this scenario, we want the happy music to have higher valence and arousal scores, and the sad music to have lower valence arousal scores. During the GPBERT model inference, the softmax operation would first calculate a score for every data to indicate its probability of having a high valence/arousal label (pre-argmax score)\footnote{There is also a score for low valence/arousal in the final layer.}, and the argmax operation would give every piece a binary classification result (post-argmax label). 
There are two metrics calculated in the table. \texttt{high valence/arousal percentage (HVP or HAP)} calculates the percentage of music having high valence/arousal post-argmax label, and \texttt{mean valence/arousal score (MVS or MAS)} calculates the mean valence/arousal score from the pre-argmax score. Note that they are all on a scale from 0.0 to 1.0, and we use 0.5 to separate high valence/arousal from low valence/arousal during inference.
In theory, happy music would have higher \texttt{high valence/arousal percentage} and \texttt{mean valence/arousal score}, and sad music would have lower scores. Therefore, we then calculate the difference of \texttt{high valence/arousal percentage} and \texttt{mean valence/arousal score} between happy music and sad music groups, and a larger difference means a better model in making music convey happiness and sadness.


\begin{table*}[h!]
\centering
\setlength{\tabcolsep}{5pt} 
\captionsetup{skip=6pt} 
\caption{Comparison between our model (MoodLoopGP) and LooperGP in happy-sad emotion score difference. HVP and HAP stand for high valence percentage and high arousal percentage, and MVS and MAS stand for mean valence score and mean arousal score.}
\begin{tabular}{l c c c c}
\toprule
\textbf{Settings} & \textbf{HVP} & \textbf{MVS} & \textbf{HAP} & \textbf{MAS} \\
\midrule
MoodLoopGP - Happy                            & 0.6573 & 0.6553 & 0.5731 & 0.5107 \\
MoodLoopGP - Sad                              & 0.2025 & 0.2165 & 0.0307 & 0.0797 \\
\textbf{MoodLoopGP - Difference}              & \textbf{0.4548} & \textbf{0.4388} & \textbf{0.5424} & \textbf{0.4310} \\
\midrule
LooperGP - Happy                         & 0.3666 & 0.3784 & 0.1414 & 0.1828 \\
LooperGP - Sad                           & 0.3425 & 0.3652 & 0.1308 & 0.1756 \\
\textbf{LooperGP - Difference}           & 0.0241 & 0.0132 & 0.0106 & 0.0072 \\
\bottomrule
\end{tabular}
\label{compare_baseline}
\end{table*}


The classification score results from epoch 20 are displayed in Table \ref{compare_baseline}, along with a comparison between our work and LooperGP, which is used as a baseline. It is important to note that in LooperGP, no control tokens were used when generating either happy or sad music. In fact, there was no difference between the two settings at all, as this was done to align with MoodLoopGP. However, there indeed exists a slight variation in the classification score between different trials, but it is smaller enough to be discarded.

When comparing the performance between models, all four metrics verify that MoodLoopGP can effectively generate music with target emotion when providing the corresponding prompt, and the metrics differences between happy and sad music generated by MoodLoopGP are up to 54\%. It also shows that the training process creates an unbiased improvement over happy and sad music, which is validated by the fact that the absolute difference between \texttt{MoodLoopGP - Happy} and \texttt{LooperGP - Happy} and the difference between \texttt{MoodLoopGP - Sad} and \texttt{LooperGP - Sad} is roughly the same. Additionally, although the metrics difference in \texttt{MoodLoopGP - Difference} group for valence and arousal are roughly the same, it seems that the valence scores are more balanced compared to arousal as nearly all the arousal scores are below 0.5.

We also conducted an ablation study to investigate the contributing factors of our approach. We took out one group of features in each trial and then compared the performance. The information is divided into three categories: 1) Emotion Labels (EL): Valence and Arousal tokens. 2) Music Psychology Features (MPF): Tempo and Mode tokens. 3) Tonal Tension (TT): Cloud Diameter, Cloud Momentum, and Tensile Strain tokens.


\begin{table*}[ht]
\centering
\setlength{\tabcolsep}{5pt} 
\captionsetup{skip=6pt} 
\caption{Ablation study results of the emotion evaluation. ``All" means the proposed model (MoodLoopGP), other settings mean all the features are added but the specified one, where EL, MPF, TT stand for Emotion Labels, Music Psychology Features, Tonal Tension, respectively.}
\begin{tabular}{l c c c c}
\toprule
\textbf{Settings} & \textbf{HVP} & \textbf{MVS} & \textbf{HAP} & \textbf{MAS} \\
\midrule
All - Happy                                     & 0.6573 & 0.6553 & 0.5731 & 0.5107 \\
All - Sad                                       & 0.2025 & 0.2165 & 0.0307 & 0.0797 \\
\textbf{All - Difference}                       & \textbf{0.4548} & \textbf{0.4388} & \textbf{0.5424} & 0.4310 \\
\midrule
Missing EL - Happy                 & 0.5900 & 0.5573 & 0.2000 & 0.2494 \\
Missing EL - Sad                   & 0.2100 & 0.2313 & 0.0200 & 0.0702 \\
\textbf{Missing EL - Difference}   & 0.3800 & 0.3260 & 0.1800 & 0.1792 \\
\midrule
Missing MPF - Happy       & 0.5232 & 0.5247 & 0.4283 & 0.4385 \\
Missing MPF - Sad         & 0.1835 & 0.2120 & 0.0444 & 0.1054 \\
\textbf{MPF - Difference}        & 0.3397 & 0.3127 & 0.3839 & 0.3331 \\
\midrule
Missing TT - Happy                  & 0.5624 & 0.5596 & 0.5726 & 0.5310 \\
Missing TT - Sad                    & 0.1242 & 0.1438 & 0.0401 & 0.0831 \\
\textbf{Missing TT - Difference}    & 0.4382 & 0.4158 & 0.5325 & \textbf{0.4479} \\
\bottomrule
\end{tabular}
\label{ablation_study}
\end{table*}


The results in Table \ref{ablation_study} demonstrate that all the features are important for achieving the best performance. When any of the features are removed, the performance drops significantly. It should be highlighted that when the Emotion labels are missing, the HAP and MAS drop by roughly 30\%, and the HVP and MVS drop the most when the Music Psychology Features are missing. This illustrates that the Emotion Labels seem to contribute more to the arousal and Music Psychology Features seem to contribute more to the valence. Additionally, tonal tension seems to contribute more to the valence than arousal, as both the HAP and MAS Happy/Sad scores between the All and Missing TT groups are roughly the same, whereas relatively large differences are obtained for the HVP and MVS Happy/Sad scores. Removing tonal tension yields the highest difference between Happy and Sad for MAS, however it is close to the difference obtained when all features are used.

\vspace{-2mm}

\subsection{Loop Extraction}

\vspace{-1mm}

Following the evaluation approach from LooperGP \cite{adkins2023loopergp}, we use the same loop extraction method to evaluate the average number of loops per generation. The same parameters are used to implement the loop extraction algorithm, including \texttt{Minimum Repetition Notes} = 4, \texttt{Minimum Repetition Beats} = 2, \texttt{Minimum Loop Bars} = 4 and \texttt{Maximum Loop Bars} = 4. A detailed explanation of the parameters can be found in \cite{adkins2023loopergp}. We compare our model with the baseline model, which is a Transformer-XL trained on the raw DadaGP dataset instead of the loop subset in order to demonstrate the effectiveness of our model's loop generation ability, and both groups are evaluated on 2000 generations of the corresponding model.


\begin{table*}[h!]
\centering
\setlength{\tabcolsep}{5pt} 
\captionsetup{skip=6pt} 
\caption{Comparison of the average number of loops per generation between MoodLoopGP and the Transformer-XL model trained in DadaGP paper.}
\begin{tabular}{l c c}
\toprule
\textbf{Model} & \textbf{Loops Found} & \textbf{Average Number of Loop} \\
\midrule
MoodLoopGP        & 757 & 0.3789 \\
Transformer-XL-DadaGP            & 522 & 0.2702 \\
\bottomrule
\end{tabular}
\label{LoopExtraction}
\end{table*}


Table \ref{LoopExtraction} shows the loop extraction evaluation result. MoodLoopGP can generate 45\% more loops than the baseline, which demonstrates the advantage of the loop extraction algorithm is successfully kept in MoodLoopGP. We also performed a Wilcoxon Signed-Rank Test to examine the difference between MoodLoopGP and the baseline model. The result (Z = 2990.0, p < 1e-40) shows that there is a significant effect of the model type on the number of loops generated.

\vspace{-2mm}

\subsection{Subjective Evaluation}

\vspace{-1mm}

To evaluate the performance of the model from the listener's perspective, we conducted a listening test to study the generated music from the following three aspects: music quality, loop coherence, and the conveyed emotions. We recruited 11 participants, 7 male and 4 female, and approximately 2/3 of them had previously received training in music theory or musical instruments.

There were 60 musical excerpts in the listening test, and they were from three groups of 20 generations with each having 10 happy excerpts and 10 sad excerpts:
\begin{itemize}
    \item Model generations prompted with all extra information: The model with all information added in the initial prompt to guide the generation. This serves as the expected model.
    \item Model generations prompted with all information but tonal tension: This is to evaluate the contributions of the bar-level features and demonstrate the benefits by leveraging multi-granular features.
    \item Human-composed music: Human-composed music is added to serve as the baseline to investigate the difference between human and machine-composed music.
\end{itemize}
All the excerpts were randomly chosen from their group and were taken from the first four bars of the music to form a loop. Each loop is repeated several times to derive the final piece. The number of repeated times was varied between pieces with different tempos to create pieces having lengths of roughly 30 seconds. The chosen 60 excerpts were also randomized to prevent order bias during the listening test.

Additionally, all the pieces were rendered from guitar pro tablatures, which do not have dynamic information. This makes the resulting music sound rigid and different from human-performed music. To address this problem, we told the listeners to only focus on the composition part of the music rather than the performing part of the music.

After listening to every excerpt, the listeners were asked to answer the following questions:
\begin{enumerate}
    \item Have you heard the music in this excerpt before? (Prior to this survey) \texttt{(Y/N)}
    \item Do you think the music is composed by a human or a machine? \texttt{(Human/Machine)}
    \item Do you like the excerpt? \texttt{(7-point Likert scale)}
    \item Does the loop in this excerpt sound coherent to you? \texttt{(7-point Likert scale from dislike to like)}
    \item What emotion do you think this excerpt conveys? \texttt{(7-point Likert scale from sad to happy)}
\end{enumerate}


The first question investigates if participants have heard the music before to evaluate biases from prior listening experiences. The second and third questions evaluate music quality based on the assumption that human-composed music and music preferred by listeners indicate higher quality. The fourth question evaluates the quality of the generated music as loops, and the fifth question evaluates it from the emotion's perspective.

\begin{table}[t]
\centering
\setlength{\tabcolsep}{5pt} 
\captionsetup{skip=6pt} 
\caption{Percentage of heard and not heard music reported by the participants.}
\begin{tabular}{l c c}
\toprule
\textbf{Composition Type} & \textbf{Heard} & \textbf{Not Heard} \\
\midrule
Machine-Composed: All Information           & 1.82\% & 98.18\% \\
Machine-Composed: Without Tonal Tension    & 2.73\% & 97.27\% \\
Human-Composed Music                        & 5.45\% & 94.55\% \\
\bottomrule
\end{tabular}
\label{heard_or_not}
\end{table}

\begin{table}[t]
\centering
\setlength{\tabcolsep}{5pt} 
\captionsetup{skip=6pt} 
\caption{Turing Test: Percentage of music identified as human-composed or machine-composed.}
\begin{tabular}{l c c}
\toprule
\textbf{Composition Type} & \textbf{Human} & \textbf{Machine} \\
\midrule
Machine-Composed: All Information           & 27.73\% & 72.27\% \\
Machine-Composed: Without Tonal Tension    & 27.27\% & 72.73\% \\
Human-Composed Music                        & 50.45\% & 49.55\% \\
\bottomrule
\end{tabular}
\label{turing_test}
\end{table}

Table \ref{heard_or_not} displays the results of the first question, showing that the participants had mostly not heard any of the three music groups prior to the experiment. Table \ref{turing_test} presents the results of the Turing test, indicating that 27\% of machine-composed music was classified as human-composed, a lower percentage than the human music group. Surprisingly, only half of the human-composed music was correctly identified, possibly because the listeners are still biased by the loss of dynamic information and the use of virtual instruments.

\begin{table*}[h!]
\centering
\setlength{\tabcolsep}{3pt} 
\captionsetup{skip=6pt} 
\caption{Results for all the Likert scale questions, including the listener's preference, loop coherence (LC), Happy Emotion Scores (HES), and Sad Emotion Scores (SES).}
\begin{tabular}{l c c c c}
\toprule
\textbf{Average Score} & \textbf{Preference} & \textbf{LC} & \textbf{HES} & \textbf{SES} \\
\midrule
Machine-Composed: All Information           & -0.3045 & 0.1591 & 0.2091 & -0.2818 \\
Machine-Composed: Without Tonal Tension    & -0.2045 & 0.1682 & -0.2909 & -0.3182 \\
Human-Composed Music                        & 0.6000 & 0.9091 & 0.7091 & -0.2636 \\
\bottomrule
\end{tabular}
\label{likert_results}
\end{table*}

\begin{figure}[h!] 
  \centering
  \includegraphics[width=0.9\textwidth]{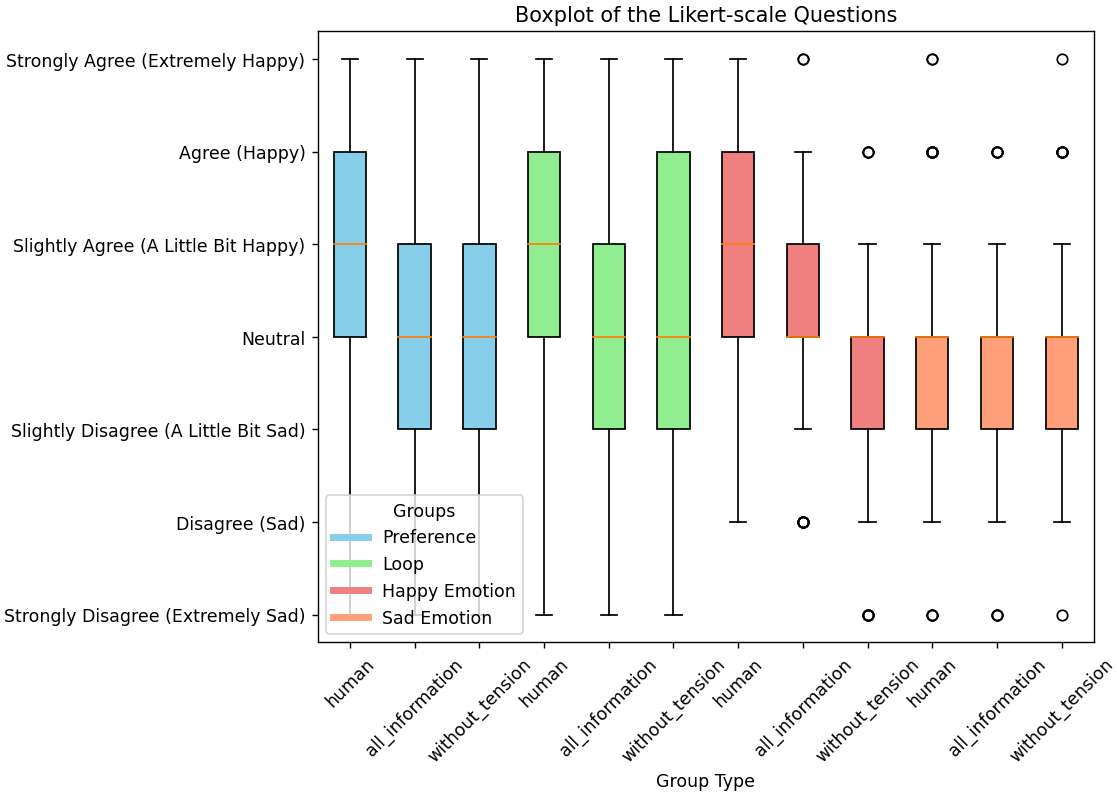} 
  \caption{Boxplot of the Likert-Scale Questions among different groups. The Happy Emotion and Sad Emotion groups correspond to pieces classified as happy and sad, respectively.}
  \label{boxplot_likert} 
\end{figure}

Table \ref {likert_results} shows the mean scores for Questions 3 to 5 on a 7-point Likert scale. The left-most answer is assigned -3, the right-most answer is assigned 3, and the stride is 1. This is to place the neutral answer (i.e., 0.) in the middle so that positive mean scores indicate positive ratings from the participants. Figure \ref{boxplot_likert} shows the boxplot of the Likert-scale questions. Human-composed music consistently outperforms machine-composed music, indicating the gaps between human-composed and machine-composed music. 
Loop coherence scores are positive but close to 0 for all generated music groups, indicating loop coherence to listeners is not very strong, and we observe a slight difference in the median loop coherence scores between human and machine groups.
The Happy Emotion and Sad Emotion Scores are obtained from results to the emotion question (Question 5) for pieces classified as happy and sad, respectively. The human group achieves the best result in HES and HES-SES difference. 
The difference obtained for the Happy and Sad pieces for the machine-composed groups (all\_information only) indicates that the generated music is successful in varying the emotional expression from sad to happy. 
The boxplot highlights that participants had difficulty differentiating happy and sad music in the \texttt{Without Tonal Tension} group, but were able to do so in the \texttt{All Information} group. Therefore, Tonal tension likely helped in generating human-perceivable happy and sad music. This is supported by the Wilcoxon Signed Rank Test results, which will be covered later.

\begin{table*}[h!]
\centering
\setlength{\tabcolsep}{5pt} 
\captionsetup{skip=6pt} 
\caption{Friedman test result for the three groups of music in the listening test.}
\begin{tabular}{l c c}
\toprule
\textbf{Question} & \textbf{$\chi^2$(2)} & \textbf{p-value} \\
\midrule
Preference                      & 51.66 & 6.06e-12 \\
Loop                            & 36.00 & 1.53e-8 \\
Emotion                         & 22.41 & 1.36e-5 \\
\bottomrule
\end{tabular}
\label{ThreeGroupKW}
\end{table*}



\begin{table*}[h!]
    \begin{minipage}{0.48\textwidth}
        \centering
        \captionsetup{skip=6pt} 
        \caption{Wilcoxon Sign Rank Test result between human-composed music and MoodLoopGP with tonal tension.}
        \begin{tabular}{l c c}
            \toprule
            \textbf{Question} & \textbf{Z} & \textbf{p-value} \\
            \midrule
            Preference & 3002.50 & 1.38e-10 \\
            Loop & 4179.50 & 4.59e-7 \\
            Emotion & 4236.50 & 8.46e-3 \\
            \bottomrule
        \end{tabular}
        \label{humanAndTensionWilcoxon}
    \end{minipage}\hfill
    \begin{minipage}{0.48\textwidth}
        \centering
        \captionsetup{skip=6pt} 
        \caption{Wilcoxon Sign Rank Test result between MoodLoopGP with and without tonal tension.}
        \begin{tabular}{l c c}
            \toprule
            \textbf{Question} & \textbf{Z} & \textbf{p-value} \\
            \midrule
            Preference & 4551.50 & 0.42 \\
            Loop & 6372.50 & 0.70 \\
            Emotion & 3860.50 & 1.11e-2 \\
            \bottomrule
        \end{tabular}
        \label{tensionOrNotWilcoxon}
    \end{minipage}
\end{table*}

In order to gain a better understanding of the outcomes of the Likert-scale questions, we conducted a Friedman test among the three groups of music. The results, as presented in Table \ref{ThreeGroupKW}, indicate that the source of generation (human, machine with all information, and machine without tension) has a significant impact on the listener's preference, loop, and emotional perspective (p<1e-4 for all three questions). 
We also carried out multiple pairwise comparisons using the Wilcoxon Sign Rank Test with  a Bonferroni-corrected $\alpha$ level ($\alpha$/3 =.0167). We found significant differences between the \texttt{Human} and \texttt{All Information} groups for preference (Z=3002.50, p<.01), loop coherence (Z=4179.50, p<.01), and emotion (Z=4236.50, p<.01), confirming that human-produced music outperforms the machine-generated one. Additionally, we were interested in exploring the effect of bar-level features (i.e., Tonal Tension) in the generation process. We did not find significant differences between the \texttt{All Information} (including tonal tension) and \texttt{Without Tonal Tension} groups for the preference (Z=4551.50, p=.42) and loop coherence (Z=6372.50, p=.70) indicating that conditioning based on tonal tension may not contribute to improving preference and loop coherence. However, we found a significant difference between the All Information and Without Tonal Tension groups (Z=3860.50, p<.0167) for emotion showing that adding tonal tension in the conditioning improves the generation of emotion-specific music.

\section{Conclusion}

\vspace{-2mm}

In this paper, we present MoodLoopGP, a novel approach for emotion-conditioned and loopable music generation utilizing multi-granular musical features. Through the integration of both song-level attributes (emotion labels, tempo, mode) and bar-level attributes (tonal tension), our model demonstrates an enhanced capacity to generate music conveying specified emotions of happiness and sadness while keeping the model's ability of music loop generation. It is supported by the empirical evaluations conducted, including algorithmic emotion classification, loop extraction, and a subjective listening test.
Our work demonstrates that incorporating music psychology features can enrich conditional generative models, and our multi-granular conditioning strategy offers a promising direction for more fine-grained control over emotion-specific music generation.

\vspace{-5mm}

\subsubsection{\ackname} This work is supported by the EPSRC UKRI Centre for Doctoral Training in Artificial Intelligence and Music (Grant no. EP/S022694/1).

\bibliographystyle{splncs04}
\bibliography{mybibliography}

\end{document}